\documentclass[aps,pra,reprint,superscriptaddress]{revtex4-2}

\usepackage{graphicx}
\usepackage{amsmath}
\usepackage{amssymb}
\usepackage{xcolor}
\definecolor{darkblue}{rgb}{0,0,.65}
\definecolor{darkgreen}{rgb}{0.2,0.6,0.2}

\usepackage[colorlinks=true,linkcolor=blue,urlcolor=blue,citecolor=blue]{hyperref}
\usepackage[normalem]{ulem} 
\definecolor{mgreen}{RGB}{1,123,0}

\begin{document}

\title{Optimal control for preparing fractional quantum Hall states in optical lattices}
	
\author{Ling-Na Wu}
\affiliation{Center for Theoretical Physics and School of Physics and Optoelectronic Engineering, Hainan University, Haikou, Hainan 570228, China}

\author{Xikun Li}
\email{xikunli@ahu.edu.cn}
\affiliation{School of Physics and Optoelectronic Engineering, Anhui University, Hefei, Anhui 230601, China}

\author{Nathan Goldman}
\affiliation{Center for Nonlinear Phenomena and Complex Systems, Université Libre de Bruxelles, CP 231, Campus Plaine, 1050 Brussels, Belgium}
\affiliation{International Solvay Institutes, 1050 Brussels, Belgium}
\affiliation{Laboratoire Kastler Brossel, Collège de France, CNRS, ENS-Université PSL, Sorbonne Université, 11 Place Marcelin Berthelot, 75005 Paris, France}
	
\author{Botao Wang}
\email{botao.wang@ulb.be}
\affiliation{Center for Nonlinear Phenomena and Complex Systems, Université Libre de Bruxelles, CP 231, Campus Plaine, 1050 Brussels, Belgium}
\affiliation{International Solvay Institutes, 1050 Brussels, Belgium}


\begin{abstract}
Preparing fractional quantum Hall~(FQH) states represents a key challenge for quantum simulators. While small Laughlin-type states have been realized by manipulating two atoms or two photons, scaling up these settings to larger ensembles stands as an impractical task using existing methods and protocols. In this work, we propose to use optimal-control methods to substantially accelerate the preparation of small Laughlin-type states, and demonstrate that the resulting protocols are also well suited to realize larger FQH states under realistic preparation times. Our schemes are specifically built on the recent optical-lattice experiment [Leonard et al., Nature (2023)], and consist in optimizing very few control parameters:~the tunneling amplitudes and linear gradients along the two directions of the lattice. We demonstrate the robustness of
our optimal-control schemes against control errors and disorder, and discuss their advantages over existing preparation methods. Our work paves the way to the efficient realization of strongly-correlated topological states in quantum-engineered systems.

\end{abstract}

\maketitle

\section{Introduction}
The fractional quantum Hall (FQH) effect, a paradigmatic strongly-correlated phenomenon in condensed matter physics, has been a beacon for researchers exploring the interplay between quantum mechanics and topology~\cite{1999Girvin,1999Stormer,2020Halperin}. The FQH states exhibit exotic properties such as topological entanglement and fractional charge excitations~\cite{2017Wen,2021Feldman}, which makes the study of FQH states of fundamental interest. Meanwhile, FQH states could also have important impacts on practical applications in quantum information technology~\cite{2008Nayak}. For example, certain FQH states with non-Abelian anyonic excitations could be used for topological quantum computation~\cite{2003Kitaev}. It becomes an exciting task to construct a well-controlled device so as to create and manipulate FQH states in a controllable manner.

Advancements in quantum engineered systems, highlighted by the successful creation of artificial gauge fields~\cite{2011Dalibard, 2014Goldman, 2017Zhang_Manipulating, 2018Aidelsburger}, have marked a significant step forward in the pursuit of strongly correlated topological states~\cite{2016Goldman, 2019Zhang_rev,2019Cooper, 2019Ozawa, 2020Browaeys}.
Recently, L\'eonard \textit{et al.}\ successfully realized a Laughlin-type FQH state with two ultracold atoms in a 4-by-4 site optical lattice by using Floquet engineering~\cite{Leonard2023}.
{Similar FQH states were also explored using rapidly rotating ultracold atoms~\cite{gemelke2010,2024Lunt}, interacting polaritons~\cite{Clark2020}, and photons in a circuit quantum electrodynamics system~\cite{2024WangCan}.}
These approaches, while groundbreaking, set the stage for further optimization and exploration of FQH states. In particular, the challenge remains to {reduce the preparation time to limit undesirable effects like Floquet heating due to periodic driving forces,} to enhance the fidelity and robustness of state preparation, especially in the presence of experimental noise and disorder. Most importantly, the need to extend such achievements to larger systems is imperative for advancing quantum simulations and quantum computing.

In this context, the work presented in Ref.~\cite{Blatz2023} employs Bayesian optimization (BO), a machine learning technique, to refine the experimental protocol for realizing the FQH state with ultracold atoms~\cite{Leonard2023}. 
Different from the experimental protocol based on following the many-body gap across the parameter space, Ref.~\cite{Blatz2023} harnesses the power of BO to search for an optimal ramping protocol. Each parameter is modeled as a piece-wise linear function, 
which is optimized to reduce the overall experimental costs. Compared to manual ramp design, the ramp protocol optimized by BO is capable of achieving equivalent fidelity (between the final state and the target state) at a speed that is ten times faster, even when considering the disorder present in realistic experimental settings.

In the realm of quantum control, the Bayesian optimization approach, as described in Ref.~\cite{Blatz2023}, is not the sole frontier; optimal control theory (OCT) stands as a powerful alternative, with its broad applicability across diverse physical systems, including nuclear magnetic resonance (NMR)~\cite{KHANEJA2005} and ultracold atoms~\cite{vanFrank2016,Li2018}, etc. OCT typically harnesses two principal classes of optimization methodologies: (i) local optimization strategies, with notable examples including Krotov~\cite{Sklarz2002}, GRAPE~\cite{KHANEJA2005}, CRAB~\cite{Doria2011}, GROUP~\cite{Sorensen2018}, and GOAT~\cite{Machnes2018}; and (ii) global optimization strategies, exemplified by differential evolution (DE)~\cite{Storn1997} and covariance matrix adaptation evolution strategy (CMA-ES)~\cite{Hansen2006}. While analytic solutions are readily at hand for scenarios where the quantum system's Hilbert space is of low dimension~\cite{Lloyd2014,Alessandro2001,Boscain2002,Khaneja2001,Boscain2005,Boscain2006,Boscain2014,Hegerfeldt2013,Hegerfeldt2014,Boozer2012,Jafarizadeh2020},  high-dimensional quantum systems demand the invocation of numerical optimization techniques. 
This numerical reliance has made the application of OCT in strongly correlated systems largely uncharted. 


In this paper, we leverage OCT to speed up the preparation of FQH states. We achieve this by employing cubic spline functions for the control parameters within the ramp protocol, which are refined using CMA-ES to enhance the fidelity of the system.
Our optimized ramp protocol has proven to be efficient, achieving higher fidelity at a faster pace compared to previous results~\cite{Blatz2023,Leonard2023}. We also demonstrate its resilience against the impact of disorder and control errors. 
Moreover, our work not only underscores the successful implementation of these optimization techniques in the existing (Harvard) 4-by-4 lattice with two bosons, but also highlights their scalability to larger system configurations~(systems of 8-by-4 lattice sites
with four bosons are explicitly addressed in this work).

The paper is organized as follows: Section II describes the theoretical background and the model that forms the basis of our investigation. Section III presents an in-depth analysis of the performance of our optimized protocols within a 4-by-4 lattice, specifically for two particles. This section accentuates the effectiveness of our methods, showcasing not only their better performance but also their resilience in the presence of noise. Building upon the findings for the small system, Section IV extends our discussion to larger system sizes, demonstrating the scalability of our approach and its continued relevance as the complexity of the system increases. Section V wraps up the paper with a summary of the main results.

\section{Theoretical Background and Model}

Considering strongly-interacting bosons moving in a two dimensional square lattice subject to a uniform magnetic flux, the system can be described by the following Harper-Hofstadter-Hubbard (HHH) Hamiltonian~\cite{Harper1955, Hofstadter1976,2005Sorensen},
\begin{align}
	\hat{H}_{\rm HHH} \!=\! 
	&- t_x \sum_{x, y} \left(e^{-i \phi y} \hat{a}^\dagger_{x+1, y} \hat{a}^{\vphantom{\dagger}}_{x, y} + \mathrm{H.c.}\right)\notag\\
	&- t_y \sum_{x, y} \left(\hat{a}^\dagger_{x, y+1} \hat{a}^{\vphantom{\dagger}}_{x, y} + \mathrm{H.c.}\right)\notag\\
	&+ \frac{U}{2} \sum_{x, y} \hat{n}_{x, y} (\hat{n}_{x, y} - 1).\label{eq:HH}
\end{align}
Here, \( \hat{a}^\dagger_{x, y} \) and \( \hat{a}_{x, y} \) are the bosonic creation and annihilation operators at lattice site \( (x, y) \), and \( \hat{n}_{x, y} \!=\!\hat{a}_{x, y}^\dagger \hat{a}_{x, y}\) counts the corresponding particle number,
$t_x$ and $t_y$ denote the strength of the nearest-neighbor tunneling along the $x$ and $y$ direction, respectively; $U$ is the on-site interaction strength. 
The Peierls phase factor $e^{-i \phi y}$ corresponds to a uniform magnetic flux $\phi\!=\!2\pi\alpha$ per plaquette.

The HHH model is a paradiagmatic model of fractional Chern insulators (FCIs), which are lattice analogues of FQH states~\cite{Sheng2011,Neupert2011,Regnault2011,2023Liu}. 
In the strongly-interacting regime, it has been numerically established that the HHH model hosts a bosonic FCI state akin to the Laughlin state at filling $\nu\!=\!1/2$~\cite{2005Sorensen,2006Palmer,2007Hafezi,2017Gerster,2019Rosson,2018Dong, 2018Raviunas, 2020Macaluso, 2020Motruk, 2020Repellin, 2012Kjaell, 2019Repellin,2021Palm_snapshot, 2021Dehghani, 2021Cian,2022Wang,2022Boesl}. The topologically nontrivial properties can be characterized by e.g.\ a fractional many-body Chern number $\nu_{\text{MB}}\!=\!1/2$~\cite{1985Niu,1986Tao}. Despite the identification of various preparation schemes~\cite{2004Popp, 2013Cooper, 2013Yao, 2014Kapit,2014Grusdt, 2015Barkeshli, 2017Repellin, 2017Motruk, 2017He, 2019Hudomal, 2020Andrade,2024Palm,2024Wang_ele}, their experimental realization in optical lattices with ultracold atoms has been challenging.

Only recently, the realization of the $\nu\!=\!1/2$ FCI state with ultracold atoms was achieved by using two atoms in a strongly-interacting optical lattice~\cite{Leonard2023}. Facilitated by the ability of site-resolved control in a quantum gas microscope, the authors designed an adiabatic-state-preparation scheme. Specifically, two linear gradient potentials were introduced in order to prepare the target FCI state. The corresponding system Hamiltonian reads
\begin{align}
	\hat{H} \!=\! 
	\hat{H}_{\rm HHH} + \Delta_x \sum_{x, y} x \, \hat{n}_{x, y} + \Delta_y \sum_{x, y} y \,\hat{n}_{x,y},\label{eq:Hamiltonian}
\end{align}
with \( \Delta_x \) and \( \Delta_y \) being the potential gradients in the \( x \) and \( y \) directions, respectively.
A ramp protocol was employed to adiabatically steer the system from an initial state with two localized particles in a $4\times 4$ lattice to the target state. This protocol involved the sequential adjustment of the four control parameters $\{t_x,t_y,\Delta_x,\Delta_y \}$. The ramping speed for each control parameter was set to be inversely proportional to the energy gap, a technique aimed at suppressing the system's excitation. Inspired by the high controllability of various system parameters, the protocol presented in Ref.~\cite{Blatz2023} involved the simultaneous adjustment of all four control parameters, following a piece-wise linear function characterized by a set of parameters, which were optimized by using BO.

In this work, we employ the CMA-ES algorithm~\cite{Hansen2006} to do the optimization.
The control fields are determined by using a cubic spline function to interpolate between a set of parameters~{evenly spaced in time}. The tuning of these parameters is directed towards increasing the fidelity between the final achieved state $|\psi_f\rangle$ and the target state $|\psi_{\rm target}\rangle$,
\begin{align}
    F\!=\!|\langle \psi_f|\psi_{\rm target}\rangle|^2.
\end{align}
The fidelity serves as a reliable performance metric, which was also utilized in the experiment~\cite{Leonard2023}, where it was assessed by reversing the ramp process and measuring the fraction of the system that returns to its initial state.
For the experimental protocol~\cite{Leonard2023}, it was measured to be $F \!=\! 43(6)\%$ over a total evolution time of $T\!=\!100\tau$, with the tunneling time \( \tau \!=\! 4.3(1) \, \mathrm{ms} \). In Ref.~\cite{Blatz2023}, BO identified an optimal ramp with $F \!=\! 94.5\%$ over the same evolution time $T\!=\!100\tau$. When the evolution time was curtailed to $T \!=\! 20 \tau$ and $T \!=\! 10 \tau$, the fidelity gravitated to $78\%$ and $53\%$, respectively. In the following, these fidelity values establish the benchmark against which the merits of our proposed schemes are evaluated.

\section{Optimization of the Harvard experiment}\label{sec:4by4}

\begin{figure}[htp]
	\centering
	\includegraphics[width=0.99\linewidth]{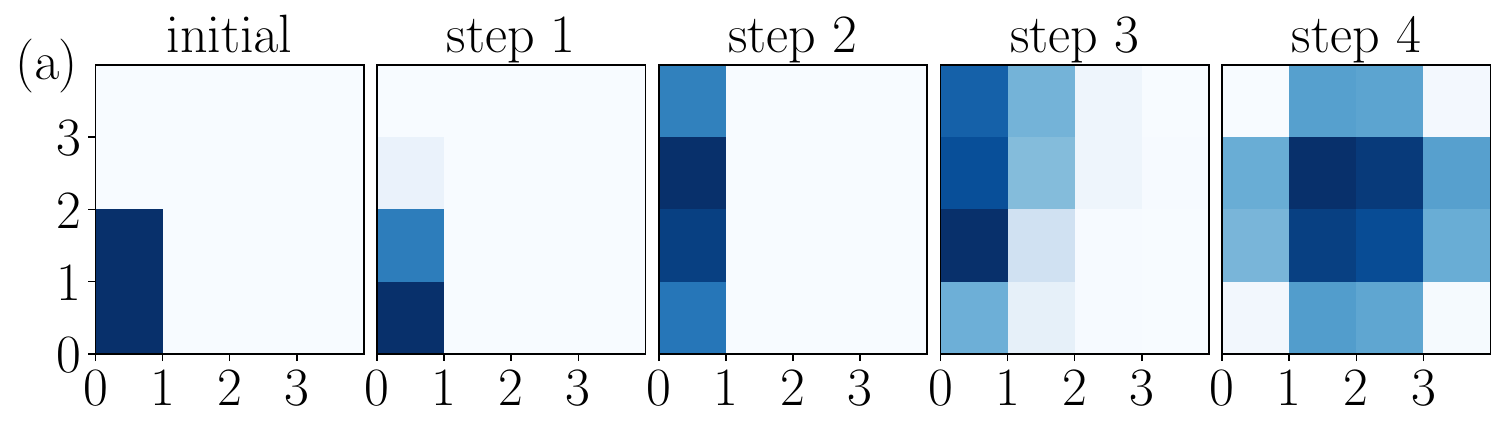}
 \includegraphics[width=0.99\linewidth]{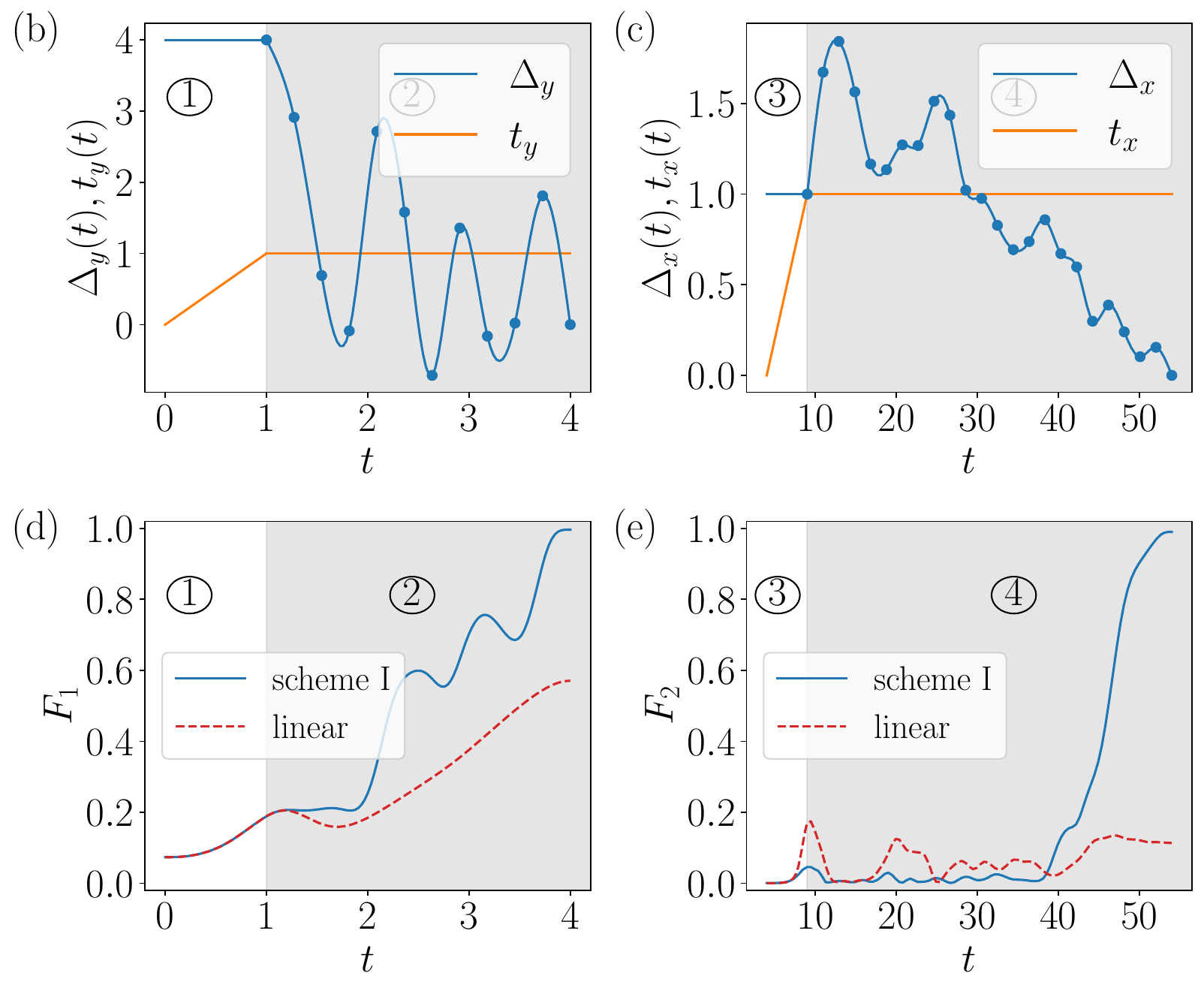}
	\caption{FQH state preparation for a $4\times 4$ lattice with two particles using Scheme I. (a) Particle density distribution during the whole process of state preparation. The first plot shows the particle density at the initial moment, where two localized atoms are on two sites. The following four plots show the particle density of the final state at each step. (b) Linear ramp $t_y(t)$ in step 1 and optimized control field for tilt $\Delta_y(t)$ in step 2~(gray shading). (c) Linear ramp $t_x(t)$ in step 3 and optimized control field for tilt $\Delta_x(t)$ in step 4~(gray shading). The blue bullets denote the interpolating parameters optimized. (d,e) Evolution of the fidelity between the instantaneous state and the corresponding target states, $F_i\!=\!|\langle \psi_{i,\rm target}|\psi(t)\rangle|^2$. 
 A nearly perfect fidelity $F\!=\!99.0\%$ with the target state in step 4, i.e., the Laughlin-type FQH state, is obtained with a duration $T_1+T_2+T_3+T_4 \!=\! 1\tau+3\tau+5\tau+45 \tau\!=\!54\tau$. Optimization is conducted utilizing the Python package CMA-ES~\cite{hansen2019pycma}. As a comparison, the linear scheme gives a low fidelity within the same time, as shown by the red dashed line.
	}\label{L4hc1}
\end{figure}

In this section, we focus on the system of $4\times 4$ lattice with two bosonic particles. We directly simulate the effective Hamiltonian instead of modeling the Floquet sequence as performed in the experiment~\cite{Leonard2023}. The state preparation begins from an initial state of two localized atoms along the $y$-axis, as shown in the first panel of Fig.~\ref{L4hc1}(a). The target state is chosen as the ground state of the Hamiltonian (\ref{eq:Hamiltonian}) with $\{t_x\!=\!\hbar/\tau,t_y\!=\!\hbar/\tau,\Delta_x\!=\!0,\Delta_y\!=\!0 \}$. We set the tunneling amplitudes in the target state, $\hbar/\tau$, as the unit of energy, and use the tunneling time $\tau$ as the unit of time. The flux is kept constant at $\phi\!=\!2\pi \times 0.26$ and the on-site interaction \( U \!=\! 8 \hbar / \tau \), so that the target state is the Laughlin-like state at filling $\nu \!=\! 1/2$~\cite{Leonard2023}. For a fair comparison with previous work~\cite{Leonard2023,Blatz2023}, we elect to navigate within an analogous parameter regime:
\begin{equation}
     0 \leq t_{x,y} \leq 1.2\hbar/\tau \quad \text{and} \quad -4\hbar/\tau  \leq \Delta_{x,y} \leq 4\hbar/\tau.\label{eq:bound}
\end{equation}
Note that the relatively small range of $t_{x,y}$ (which corresponds to small driving amplitudes in experiments) could facilitate to reduce Floquet heating caused by excitations to higher bands~\cite{Eckardt2016Interband,Blatz2023,Leonard2023}.
In the following, we employ two schemes to speed up the process of state preparation. 

\subsection{Scheme I: four-step state preparation}\label{sec:scheme1}

We first study a four-step scheme where only one parameter is tuned at each step, like the protocol applied in the experiment~\cite{Leonard2023}. As the benefits from fine-tuning the hopping amplitudes $t_{x,y}$ are rather modest~\cite{Blatz2023}, we shall linearly ramp up hopping amplitudes $t_{x,y}$ and perform optimization on the tilt potential gradients $\Delta_{x,y}$. Choosing the initial and final values of the control fields according to the experimental protocol~\cite{Leonard2023}, the state preparation consists of the following four steps:   

\begin{enumerate}
	\item Linearly ramp up $t_y$ from $0$ to $\hbar/\tau$ over duration $T_1\!=\!\tau$, and keep the other three parameters constant: $t_x\!=\!0$, $\Delta_x\!=\!\hbar/\tau$, and $\Delta_y\!=\!4\hbar/\tau$. The final state in this step is the initial state of next step. 
	
	\item While keeping $t_x\!=\!0$, $t_y\!=\!\hbar/\tau$, $\Delta_x\!=\!\hbar/\tau$ constant, optimize the ramp protocol of $\Delta_y (t)$ within time $T_2$ to maximize the fidelity between the evolved state $|\psi(T_1+T_2)\rangle$ and the intermediate target state $|\psi_{1,\rm target}\rangle$, which is the ground state of the Hamiltonian (\ref{eq:Hamiltonian}) with no tilt in $y$-direction $\Delta_y\!=\!0$. 
	
	\item Linearly ramp up $t_x$ from $0$ to $\hbar/\tau$ over duration $T_3\!=\!5\tau$, and keep the other three parameters constant: $t_y\!=\!\hbar/\tau$, $\Delta_x\!=\!\hbar/\tau$, and $\Delta_y\!=\!0$. 
	
	\item  Keeping the other three constant: $t_x\!=\!t_y\!=\!\hbar/\tau$, $\Delta_y\!=\!0$, optimize the ramp protocol of $\Delta_x (t)$ within $T_4$ to maximize the fidelity between the final state and the target FCI state $|\psi_{2,\rm target}\rangle$, which is the ground state of Hamiltonian (\ref{eq:Hamiltonian}) with tunneling strengths $t_{x}\!=\!t_y\!=\!\hbar/\tau$ and $\Delta_{x}\!=\!\Delta_{y}\!=\!0$. 
	
\end{enumerate}
The resulting density distributions for each step are depicted in Fig.~\ref{L4hc1}(a). One can see that after ramping down the tilt in $y$-direction~(step 2), the particles delocalize along this direction. The dispersion across the $x$-direction is achieved once the tilt in that direction is similarly diminished~(step 4), culminating in a flat density distribution in the bulk~[see the last panel of Fig.~\ref{L4hc1}(a)], a hallmark characteristic of the FQH state. Figures \ref{L4hc1}(b) and (c) show the control fields associated with steps 1-2 and 3-4, respectively. The optimized control fields are distinctly marked with a gray background for ease of identification. The progression of the fidelity, comparing the time evolved state to the target state of step 2 and step 4, is displayed in Figs. \ref{L4hc1}(d) and (e), respectively. 
An elevated fidelity level of $99\%$ is achieved within a total time duration of \( T \!=\! T_1 + T_2 + T_3 + T_4 \!=\! 1 \tau+ 3\tau + 5\tau + 45\tau \!=\! 54\tau \). We have also used GRAPE to perform the optimization, which gives similar results, as shown in Appendix~\ref{appendix:GRAPE}. Note that a linear ramping protocol~(red dashed lines) yields a comparatively lower fidelity within the same evolution time. Moreover, our results exceed those in Ref.~\cite{Blatz2023}, which attained a fidelity of 94.5\% at $T\!=\!100\tau$.

\begin{figure}[!htp]
	\centering
    \includegraphics[width=0.99\columnwidth]{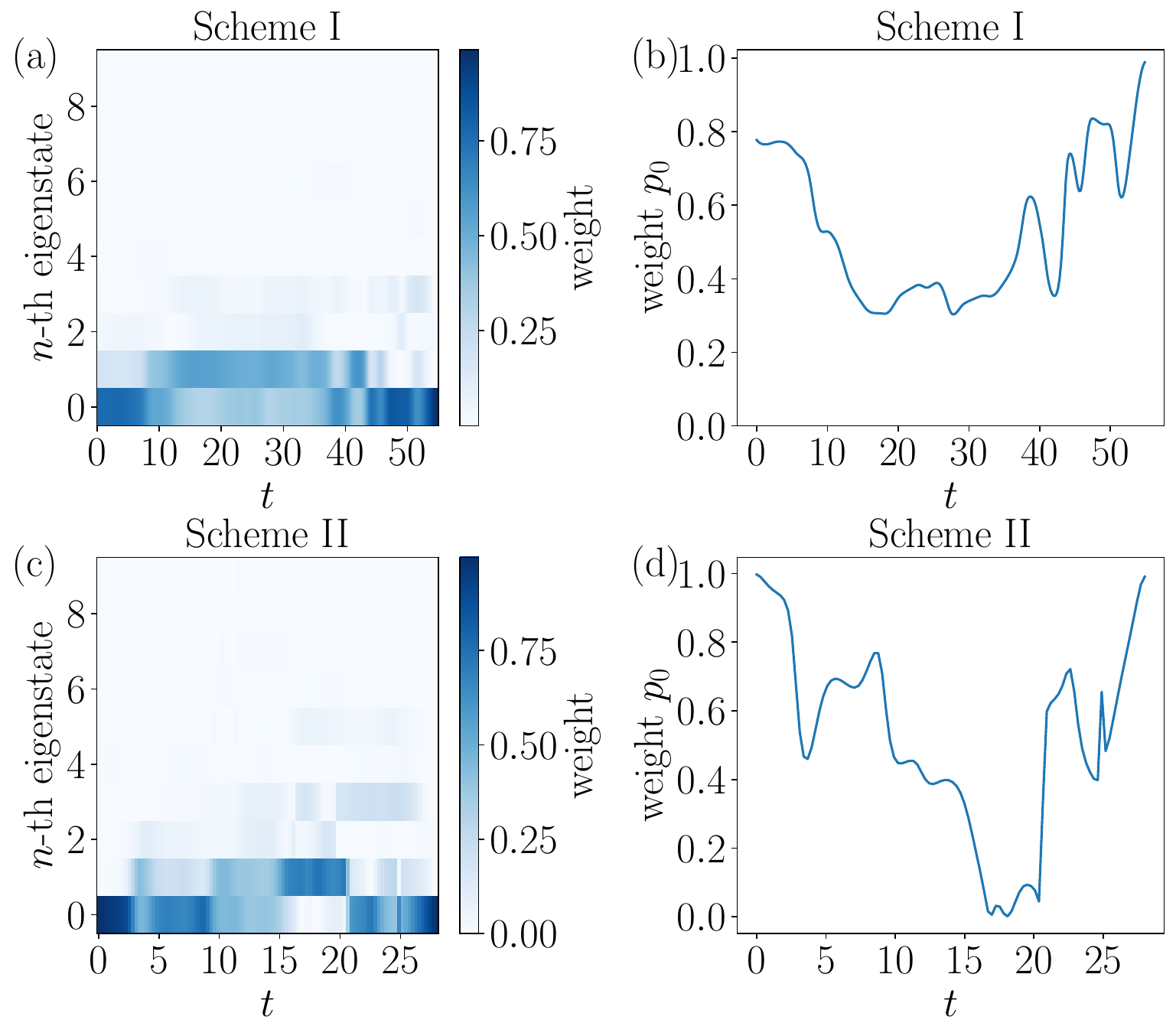}\\
	\caption{ (a,c) The weight in the lowest few eigenstates during the last step preparation for the two schemes. (b,d) The weight in the instantaneous ground state $p_0$ during the last step preparation for the two schemes.
	}\label{weight}
\end{figure}

Figures.~\ref{weight}(a,b) show the distribution of weight among the eigenstates of the instantaneous Hamiltonian during the final stage of preparation, which offers a glimpse into the excitation patterns. It is noticeable that minor excitations to the low-energy excited states occur throughout the process. Nonetheless, these excited populations eventually revert to the ground state by the end of the ramp protocol.
This observed excitation behavior is starkly distinct from that encountered in adiabatic protocols, where the parameters must be ramped very gradually—a requirement that necessitates a significantly longer evolution time—to ensure that the system remains in the ground state of the instantaneous Hamiltonian throughout the process.

%
%

\subsection{Scheme II: two-step state preparation}

The second scheme involves a two-step process: (i) the tunneling parameter $t_y$ and tilt $\Delta_y$ along the $y$-axis are optimized simultaneously to prepare the ground state of Hamiltonian~\eqref{eq:Hamiltonian} with $t_y\!=\!\hbar/\tau$, $\Delta_y\!=\!0$, $t_x\!=\!0$, and $\Delta_x\!=\!100\hbar/\tau$~(a large value of $\Delta_x$ lifts the degeneracy along $x$-axis); (ii) the tunneling parameter $t_x$ and tilt $\Delta_x$ along the $x$-axis are optimized to prepare the ground state of Hamiltonian~\eqref{eq:HH} with $t_x\!=\!t_y\!=\!\hbar/\tau$.

\begin{figure}[!htp]
	\centering
    \includegraphics[width=0.99\columnwidth]{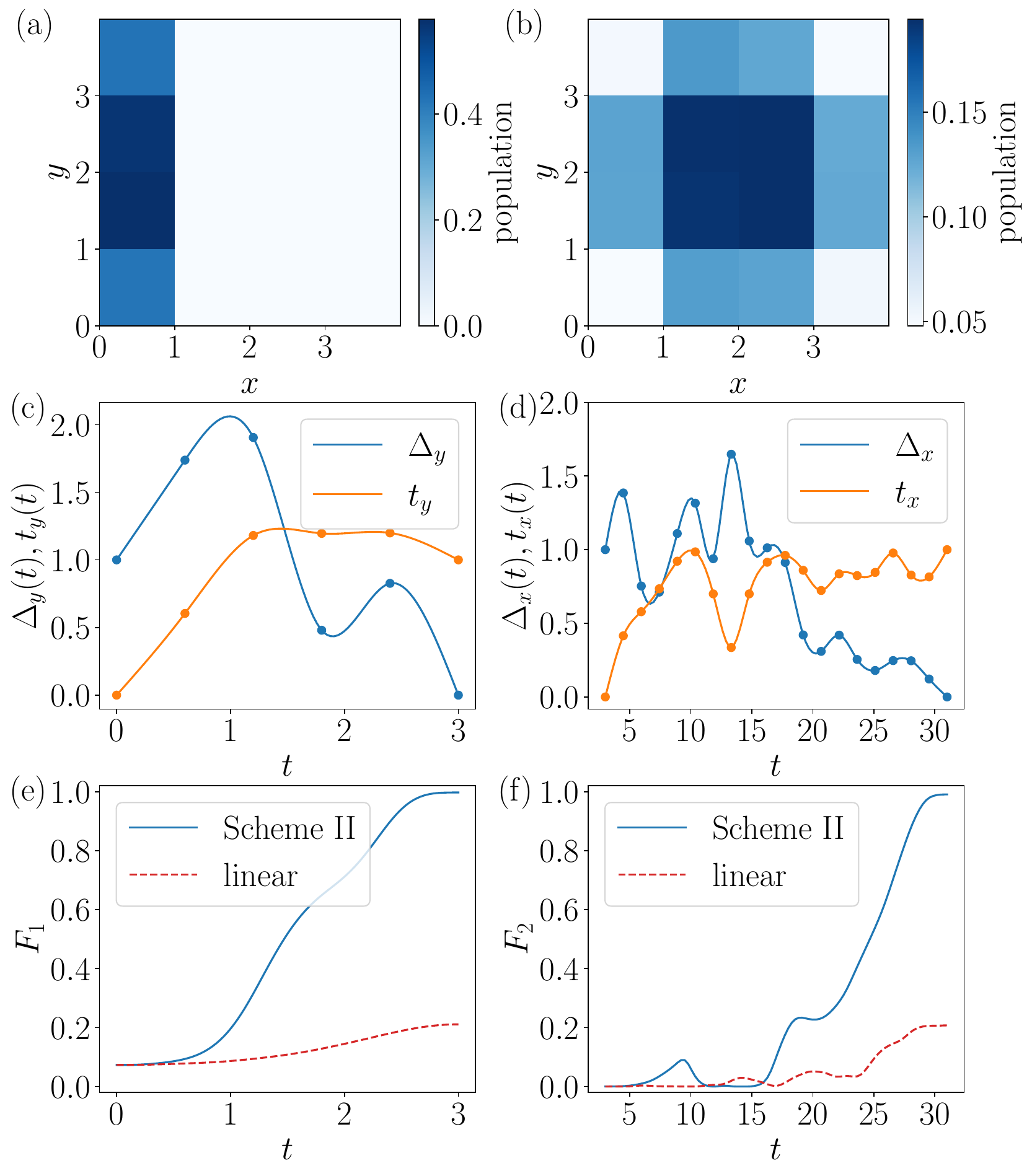}\\
	\caption{Particle density distribution of the final state for the first step (a) and the second step (b) of Scheme II. (c,d) The optimized control fields for  Scheme II. (e,f) The evolution of the fidelity between the instantaneous state and the corresponding target state, $F_i\!=\!|\langle \psi_{i,\rm target}|\psi(t)\rangle|^2$. The optimized parameters are represented by dots, except for the ends. The optimized parameters are interpolated using a cubic spline to obtain the control field. Optimization is conducted utilizing the Python package CMA-ES~\cite{hansen2019pycma}.
	}\label{L4N2_hc}
\end{figure}

Figures~\ref{L4N2_hc}(a,b) show the density distributions for the final evolved states of these two steps, and
Figs.~\ref{L4N2_hc}(c,d) showcase the optimized control fields for the two steps, respectively. The evolution of fidelity between the time evolved state and the corresponding target state in each step is depicted in (e,f). Notably, a fidelity of $99.5\%$ is achieved within $T_1 \!=\! 3 \tau$ for the first step, while for the second step, a fidelity of $99\%$ is attained within $T_2 \!=\! 28 \tau$. To summarize, a final fidelity of $99\%$ is reached with a total evolution time of $T\!=\!T_1+T_2\!=\!31\tau$. 
This outcome surpasses the performance of Scheme I. The populations of excited states are illustrated in Figs.~\ref{weight}(c,d), revealing a similar pattern as that observed in Scheme I.

\begin{figure}[!htp]
	\centering    \includegraphics[width=0.8\columnwidth]{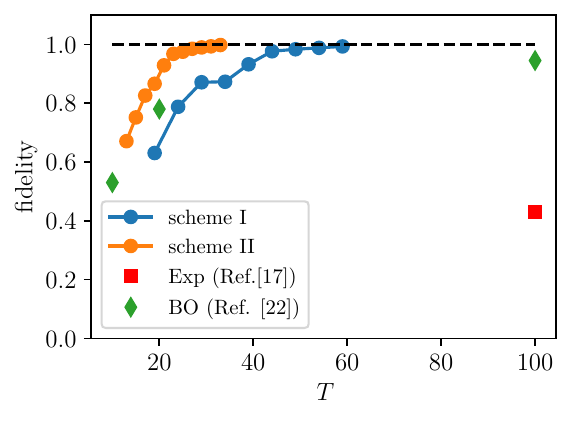}\\
\caption{The fidelity as a function of the total evolution time. The blue~(orange) bullets denote the results for Scheme I (II). Since the final stage of the process dominates the time cost, we fix the evolution time of the preceding steps and vary only the duration of the last step for both schemes. The red squares and green diamonds denote the results of the experimental protocol~\cite{Leonard2023} and the optimized approach in Ref.~\cite{Blatz2023}, respectively.
	}\label{fid_compare}
\end{figure}

Figure~\ref{fid_compare} compares the fidelity performance of our schemes  with existing methods, demonstrating the superior efficiency of our schemes.
While the experimental protocol in Ref.~\cite{Leonard2023} achieves a fidelity of {43(6)\%} at a total evolution time of \( T = 100\tau \) (red squares), and the optimized approach in Ref.~\cite{Blatz2023} reaches {94.5\% fidelity} at the same evolution time (green diamonds), 
our schemes achieve higher fidelity within substantially reduced evolution time.
Specifically, Scheme I (blue bullets) surpasses {95\% fidelity} for \( T > 44\tau \).
Scheme II (orange bullets) reaches the same high-fidelity threshold ({\(>95\%\)}) in even less time, requiring only \( T > 23\tau \).
Even when compared to the shorter evolution times in Ref.~\cite{Blatz2023}—where the fidelity drops to {78\% at \( T = 20\tau \)} and {53\% at \( T = 10\tau \)}—our schemes maintain robust performance at comparable or shorter durations.

\subsection{Robustness to noise}

\subsubsection{Control error}
In experiments, achieving the desired control field with perfect precision is often challenging. To evaluate the resilience of our protocols against potential inaccuracies in the control fields,
we subject the optimized control fields to random perturbations. Specifically, the control inaccuracies are modeled as white noise that is added onto the control field. 
This white-noise assumption aligns with the dominant noise characteristics in cold-atom experiments.
The noise, represented by \(\delta(t)\), is applied with a frequency of \(f\). The perturbed control field at time \(n\Delta t\) (where \(n\) is an integer and \(\Delta t \!=\! 1/f\)) is expressed as \(h_x(n\Delta t) \!=\! h_x^{\mathrm{Opt}}(n\Delta t) + \delta(n\Delta t)\). Here, \(h_x^{\mathrm{Opt}}(t)\) signifies the ideal control field, and the noise \(\delta (t)\) is a 
random variable drawn from a normal distribution with mean \(\mu \!=\! 0\) and standard deviation \(\sigma\). The final perturbed control field at arbitrary time is obtained by interpolating these values using a cubic spline function.

\begin{figure}[!htp]
	\centering    \includegraphics[width=0.99\columnwidth]{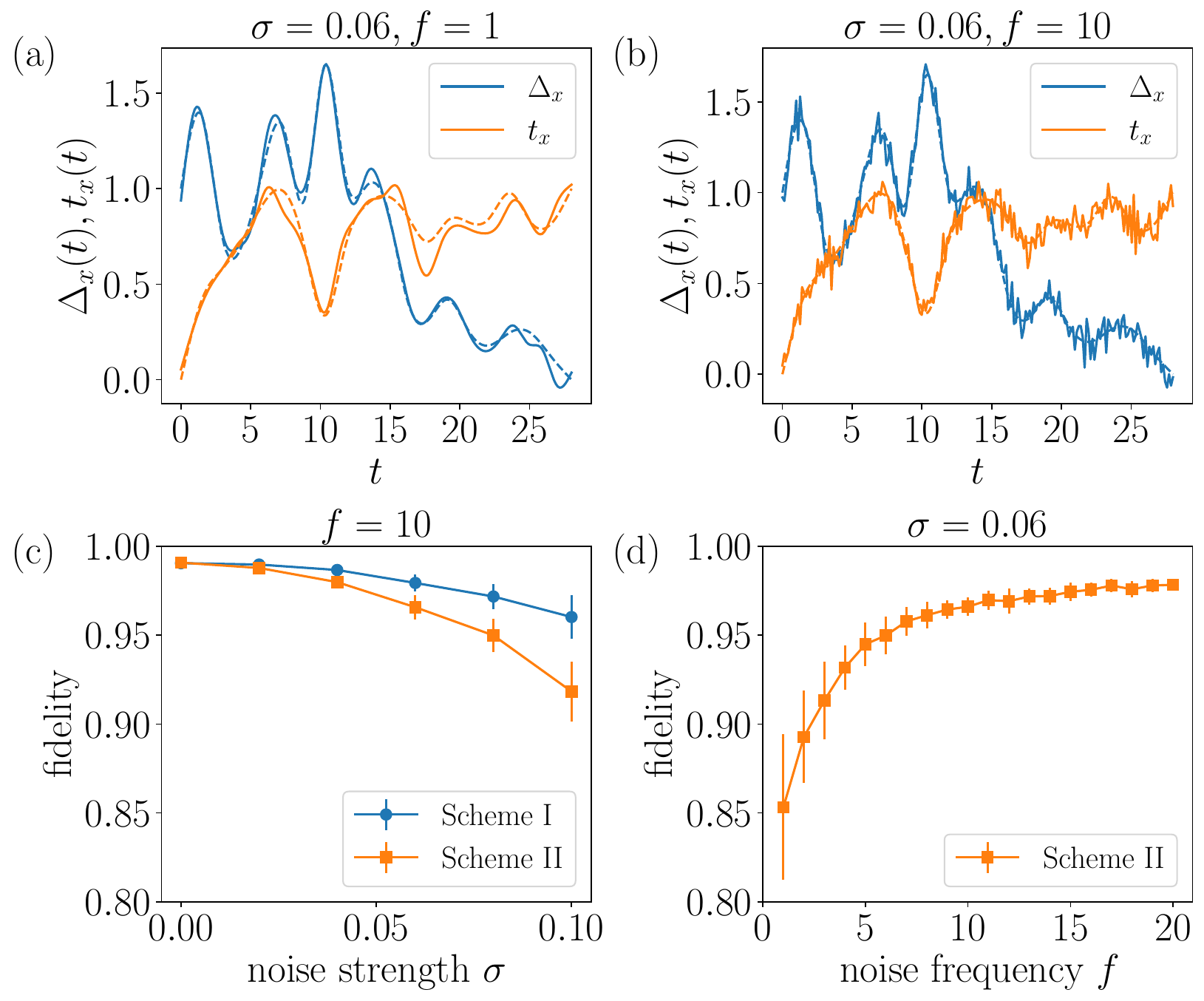}\\
	\caption{(a,b) Example noisy control profiles at $\sigma\!=\!0.06\hbar/\tau$ for frequencies $1/\tau$ and $10/\tau$, respectively; Dashed lines denote the ideal profiles.
    (c) Fidelity versus noise strength at $f\!=\!10/\tau$; 
    (d) Frequency-dependent fidelity at fixed $\sigma\!=\!0.06\hbar/\tau$.  The results are obtained by averaging over $50$ runs. The errorbars denote one standard deviation. The parameters are $L_x\!=\!L_y\!=\!4$, $N\!=\!2$, $\phi\!=\!2\pi \times 0.26$, $U\!=\!8\hbar/\tau$. 
	}\label{noise4}
\end{figure}

A depiction of the noisy control fields (with noise strength $\sigma\!=\!0.06\hbar/\tau$)  for step 2 of Scheme II is provided in Fig. \ref{noise4}(a) for  noise sampling frequency $f\!=\!1/\tau$ and in Fig. \ref{noise4}(b) for $f\!=\!10/\tau$. We conduct the time evolution with the perturbed control fields across 50 iterations and determine the average fidelity, which is then plotted as a function of the noise strength $\sigma$ in Fig.~\ref{noise4}(c). Both schemes exhibit obvious robustness against control noise. Notably, Scheme I demonstrates greater robustness compared to Scheme II, a difference we attribute to the reduced number of optimized control fields in Scheme I—only two, as opposed to four in Scheme II.
Figure~\ref{noise4}(d) quantifies the fidelity for Scheme II as a function of the noise sampling frequency, revealing sustained high performance across a broad frequency range, with particular resilience at higher frequencies. This frequency resilience may be attributed to the high-frequency noise suppression by the control field’s inherent low-pass filtering, as dictated by the cubic spline parameterization. 
This behavior is consistent with the time-correlated noise analysis (see Appendix~\ref{app:OU}) where the fidelity reduction at long correlation times $\tau_c$ follows a similar frequency-selective pattern—both cases confirm the control's high-frequency suppression capability and its inherent vulnerability to low-frequency noise components.

\subsubsection{Disorder}

\begin{figure}[!htp]
	\centering    \includegraphics[width=0.8\columnwidth]{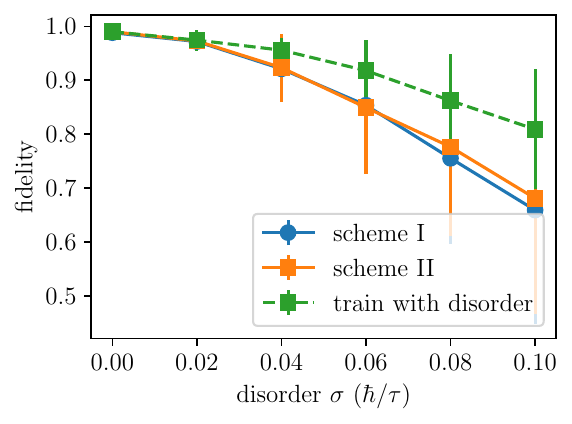}\\
	\caption{The fidelity as a function of the disorder strength. The solid lines represent the outcomes when the optimal control field, trained in the absence of disorder, is implemented. In contrast, the dashed lines illustrate the outcomes for scenarios where disorder is considered during the optimization process for Scheme II. For this latter scenario, the cost function value is derived from averaging the fidelity across 100 distinct random disorder configurations. The results for both cases are compiled by averaging over 500 instances of disorder. The error bars indicate the standard deviation from the mean. The parameters are $L_x\!=\!L_y\!=\!4$, $N\!=\!2$, $\phi\!=\!2\pi \times 0.26$, $U\!=\!8\hbar/\tau$.
	}\label{disorder4}
\end{figure}

Considering that disorder is ubiquitous in experiments, here we also evaluate the robustness of the optimized protocols against disorder. We simulate experimental conditions by including static on-site potential shifts
in our simulations. Specifically, each lattice site is impacted by a random offset. The lattice edges experience an additional, uniform offset with the corners receiving twice the impact~(caused by the creation of lattice walls via the digital micromirror devices)~\cite{Blatz2023,Leonard2023}. The (random and uniform) offsets are modeled to follow a Gaussian distribution, with an average value of $0$ and a standard deviation of $\sigma$. In Fig.~\ref{disorder4}, the fidelity of our protocols is shown as a function of the disorder strength $\sigma$. 
The mean fidelity and its standard deviation are obtained by averaging the results over 500 such disorder realizations. One can see that both protocols exhibit a similar level of endurance against the effects of disorder.

Integrating disorder into the optimization process allows for a reduction in its detrimental effects. For optimization in the presence of disorder, we use the average fidelity of 100 disorder realizations as the cost function to reduce fluctuations. The outcomes for Scheme II, represented by green squares, unequivocally demonstrate a diminished influence from disorder. This advantage intensifies as the disorder strength increases.

\section{Beyond two particles}\label{sec:6by6}

\begin{figure}[!htp]
	\centering
    \includegraphics[width=0.99\columnwidth]{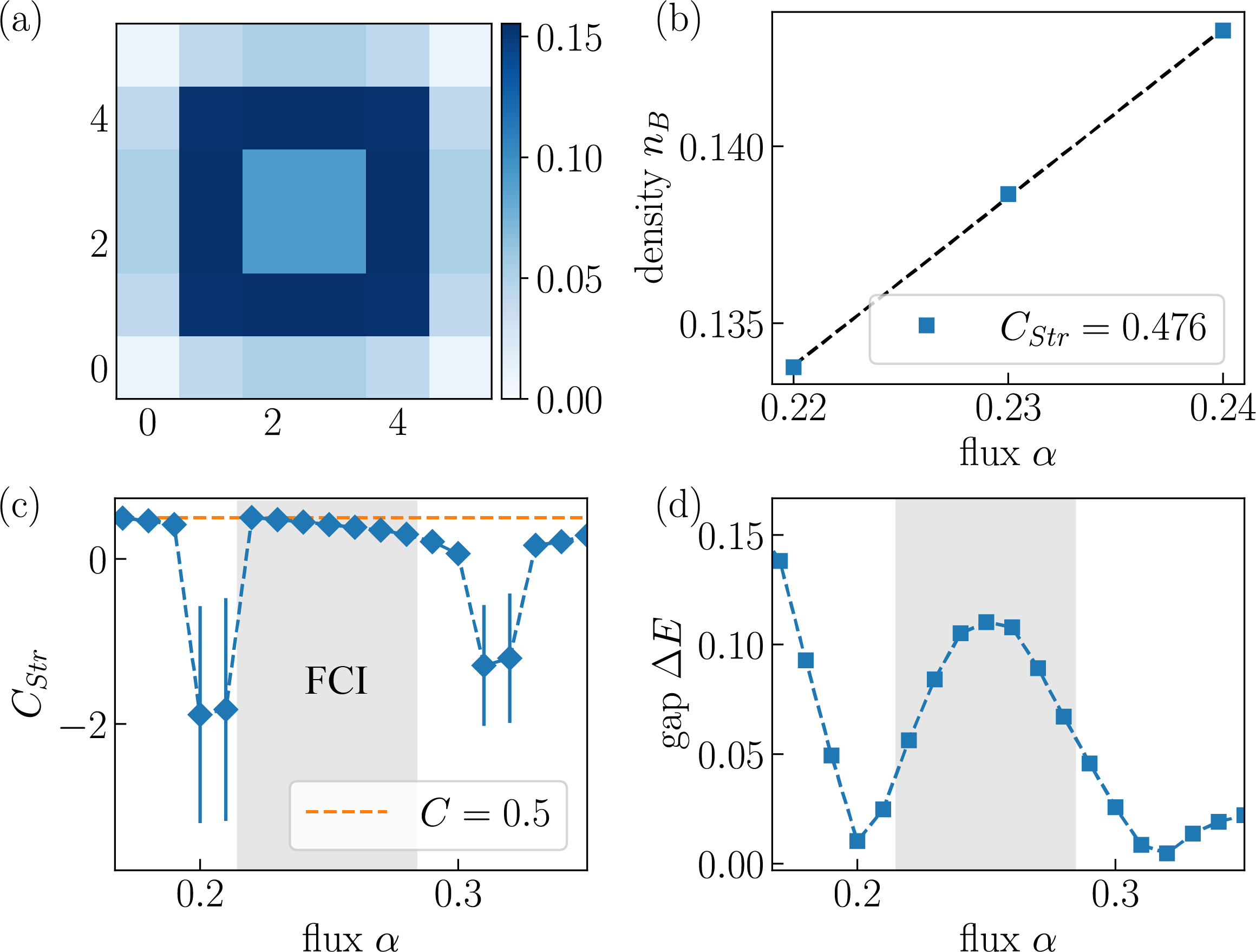}\\
	\caption{(a) The spatial density distribution of the ground state at flux $\alpha\!=\!0.22$ in the $6\times6$ system with $N\!=\!3$ particles at $U\!=\!7\hbar/\tau$. (b) The bulk density $n_B$ as a function of flux $\alpha$. (c) The Streda's marker as a function of flux $\alpha$. The expected value of \(C_{\text{Str}}\!=\!1/2\) is marked by the orange dashed line. (d) The many-body gap as a function of flux $\alpha$.
	}\label{fig_streda}
\end{figure}

Going beyond two particles, we first
focus on a system of $6\times 6$ lattice sites with $N\!=\!3$ particles. To determine the appropriate parameter regime where the system manifests properties of a fractional Chern insulator, we utilize St\v{r}eda's formula~\cite{1982Widom,1982Streda,1983Streda,2008Umucalilar,2020Repellin}, which encodes the Hall conductivity \(\sigma_{\text{H}}\) through the static spatial density distribution. The formula is given by:
\begin{equation}
    C_{\text{Str}}\!=\!\frac{\partial n_{\text{B}}}{\partial\alpha}\!=\!\frac{\sigma_{\text{H}}}{\sigma_{0}},\label{eq_Streda}
\end{equation}
where \(\sigma_0 \!=\! 1/2\pi\) represents the conductivity quantum. For the \(\nu \!=\! 1/2\) Laughlin state, \(C_{\text{Str}}\!=\!1/2\) is expected. This value serves as a benchmark for identifying the correct parameter settings. For each value of flux \(\alpha\) (in unit of \(2\pi\)), we calculate the system's ground state and determine the density distribution. An illustration of the density distribution for the ground state at \(\alpha \!=\! 0.22\) is depicted in Fig.~\ref{fig_streda}(a). The bulk density \(n_B\) within a central disk of radius \(r \!=\! 2\) is then extracted to compute \(C_{\text{Str}}\) using Eq.~\eqref{eq_Streda}. Figure~\ref{fig_streda}(b) displays the bulk density \(n_B\) as a function of \(\alpha\) (in the vicinity of \(\alpha \!=\! 0.22\)). By performing a linear fitting to the data, one can extract the St\v{r}eda's marker \(C_{\text{Str}}\) from the slope. For instance, from the data in Fig.~\ref{fig_streda}(b), \(C_{\text{Str}}\) is found to be $0.476$ for \(\alpha \!=\! 0.22\). Following this method, one can obtain \(C_{\text{Str}}\) as a function of \(\alpha\), as shown in Fig.~\ref{fig_streda}(c). The expected value of \(C_{\text{Str}}\!=\!1/2\) is marked by the orange dashed line. One can see that for a wide range of \(\alpha\), \(C_{\text{Str}}\) is close to the expected value, suggesting the existence of a fractional Chern insulating state. The deviation around \(\alpha \!=\! 0.2\) is attributed to the closing of the many-body gap, as shown in Fig.~\ref{fig_streda}(d), signifying a topological phase transition. In the following, we choose the ground state at flux \(\alpha \!=\! 0.22\) as our target state.

\begin{figure}[!htp]
	\centering
    \includegraphics[width=0.99\columnwidth]{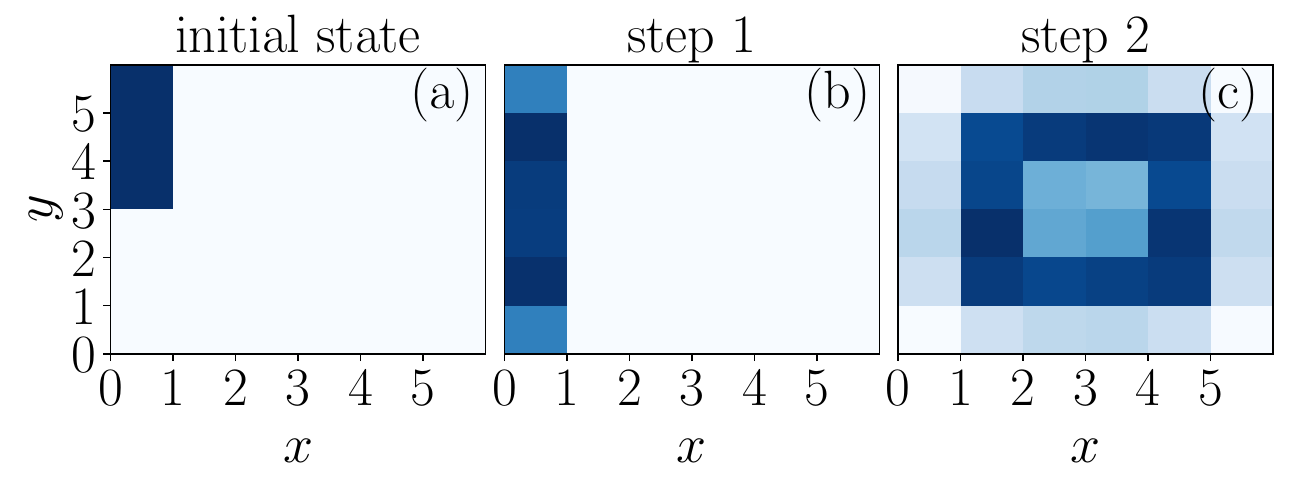}\\
    \includegraphics[width=0.99\columnwidth]{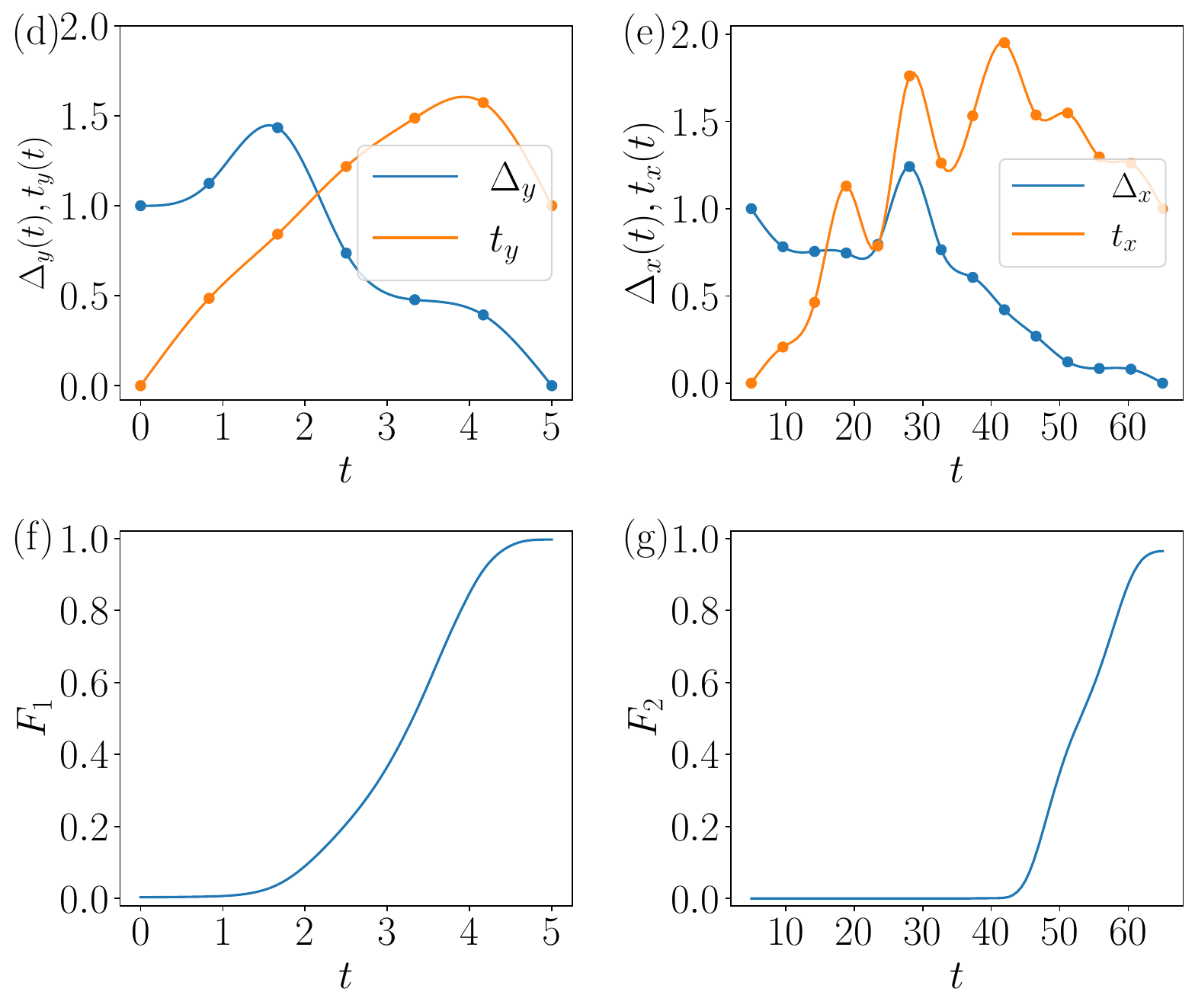}\\
	\caption{The on-site population for (a) the initial state,  (b) the final state of the first step evolution, (c) the final state of the second step evolution for the two-step scheme. (d,e) The optimized control fields for two-step scheme. (f,g) The evolution of the fidelity between the instantaneous state and the corresponding target state. The parameters optimized are represented by dots, except for the ends. The optimized parameters are interpolated using a cubic spline to obtain the control field. Optimization is conducted utilizing the Python package CMA-ES~\cite{hansen2019pycma}. The parameters are $L_x\!=\!L_y\!=\!6$, $N\!=\!3$, $\phi\!=\!2\pi \times 0.22$, $U\!=\!7\hbar/\tau$. A final fidelity of $96.4\%$ is obtained within a total evolution time of $T\!=\!5\tau+60\tau\!=\!65\tau$.
	}\label{L6N3_hc}
\end{figure}

The state preparation begins from an initial state of
three localized atoms in the $y$-axis~[see Fig.~\ref{L6N3_hc}(a)]. By means of CMA-ES strategy, we apply the two-step scheme~(scheme II) in the process of state preparation.
Figures~\ref{L6N3_hc}(b,c) show the density distribution for the evolved states at the end of these two steps, and
Figs.~\ref{L6N3_hc} (d,e) showcase the optimized control fields for the two steps, respectively. The evolution of fidelity between the evolved state and the corresponding target state is depicted in Figs.~\ref{L6N3_hc} (f,g). We find that a fidelity of $99.7\%$ is achieved within $T_1 \!=\! 5\tau$ for the first step, while for the second step, a fidelity of $96.4\%~(99.4\%)$ is attained within $T_2 \!=\! 60\tau~(100\tau)$. In the end, a fidelity of $96.4\%$ is reached with an overall evolution time of $T\!=\!T_1+T_2\!=\!65\tau$.

\begin{figure}[!htp]
	\centering
     \includegraphics[width=0.99\columnwidth]{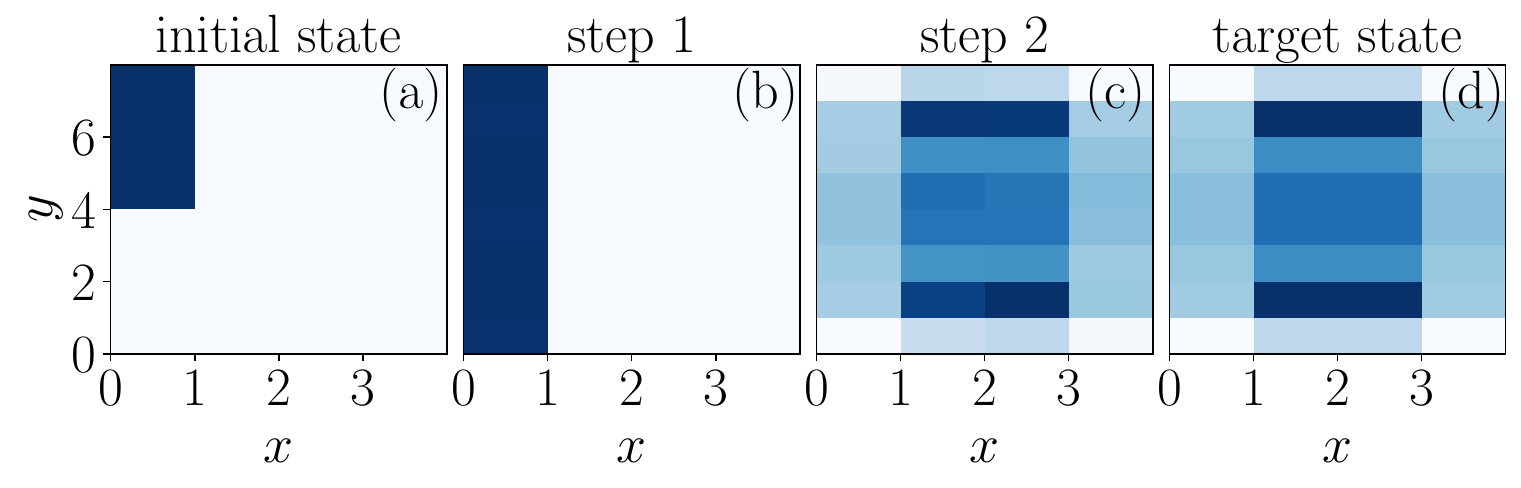}\\
    \includegraphics[width=0.99\columnwidth]{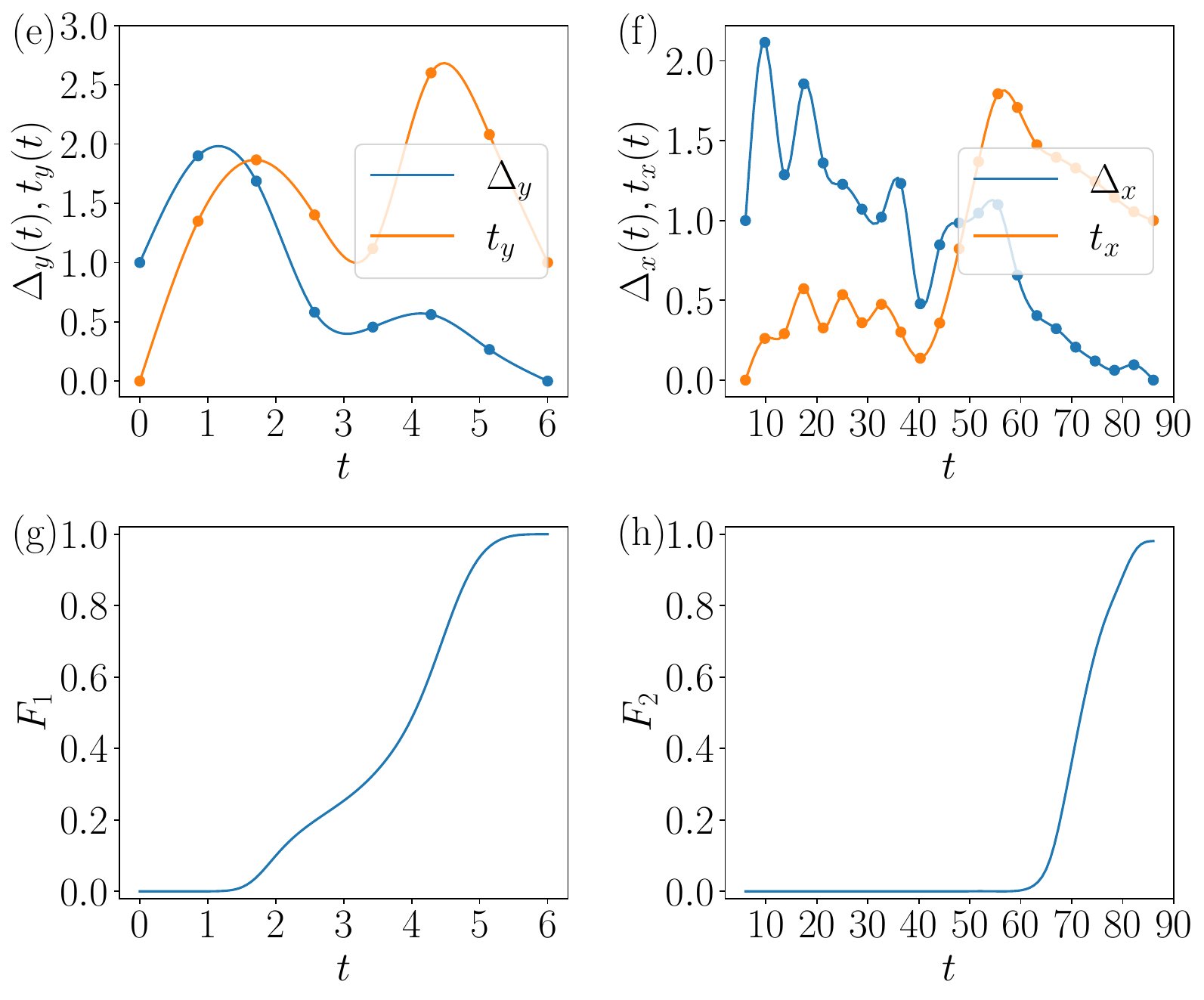}\\
	\caption{The on-site population for (a) the initial state, (b) the final state of the first step evolution, (c) the final state of the second step evolution for the two-step scheme, and (d) the target state. (e,f) The optimized control fields for two-step scheme. (g,h) The evolution of the fidelity between the instantaneous state and the corresponding target state. The parameters optimized are represented by dots, except for the ends. The optimized parameters are interpolated using a cubic spline to obtain the control field. Optimization is conducted utilizing the Python package CMA-ES~\cite{hansen2019pycma}. The parameters are $L_x\!=\!4$, $L_y\!=\!8$, $N\!=\!4$, $\phi\!=\!2\pi \times 0.3$, Hard-Core limit. A final fidelity of $98.1\%$ is obtained within a total evolution time of $T\!=\!6\tau+80\tau\!=\!86\tau$.
	}\label{L8N4_hc}
\end{figure}

Finally, we study a system of size $L_x\!=\!4$, $L_y\!=\!8$ with $N\!=\!4$ particles. In order to simplify the numerical calculations, we consider the hard-core limit ($U \rightarrow \infty$). Similar analysis based on the many-body gap and  St\v{r}eda's formula suggests that the system exhibits FCI properties~\cite{2024Palm} at flux $\alpha\!\approx\!0.3$. We employ the two-step scheme to speed up the process of state preparation. The state preparation begins from an initial state of
four localized atoms in the $y$-axis~[Fig.~\ref{L8N4_hc}(a)]. As shown in Figs.~\ref{L8N4_hc}(g,h), a final fidelity of $98.1\%$ is obtained within a total evolution time of $T\!=\!6\tau+80\tau\!=\!86\tau$.


\section{Conclusion}\label{sec:conclusions}

To sum up, we have discussed how to speed up the preparation of strongly correlated FQH states in optical lattices with ultracold atoms by means of optimal control theory. By employing cubic spline functions for control parameters and refining them using the CMA-ES algorithm, we have investigated two schemes. The first one refines the experimental protocol by optimizing the ramping speed of control parameters.
The second scheme introduces a dual-parameter tuning approach, which promises greater efficiency. Both of our optimized ramp protocols have demonstrated better performance over previous methods, achieving higher fidelity in shorter time, and showcase an impressive ability to withstand the effects of disorder and control errors.

Furthermore, our schemes have proven to be not only effective in the intensively studied 4-by-4 system with two particles, but also readily scalable to more extensive configurations. As an illustration, we have applied our methods to the preparation of FQH states in a 6-by-6 system with three particles and a 4-by-8 system with four particles. In each of these cases, our schemes have delivered good results, i.e.\ high fidelity ($\sim 98\%$) within realistic preparation time ($\sim 80\tau$). A complete characterization of the scaling behavior for arbitrarily large systems remains numerically challenging and will be an interesting open question for future investigation. Our work provides new possibility of preparing and manipulating strongly-correlated states in optical lattices with ultracold atoms, and may find broader application in quantum many body systems of increasing complexity.

 Considering realistic models that include the periodic driving, it has been shown that the Floquet heating could be suppressed in systems of few particles~\cite{2019Hudomal,2020Sun}. In general, it would be interesting to apply our optimal-control protocols in that context to improve the performance of periodically-driven many body system. Considering that optimal control protocols often exploit excited states as intermediate steps (see Fig.\ref{weight}), which is very different from typical adiabatic preparation schemes, it would be relevant to study efficient optimal preparation sequences that make use of higher bands beyond the tight-binding regime.

\begin{acknowledgments}
{We thank Brice Bakkali-Hassani for his insightful comments on our manuscript. We thank  Joyce Kwan, Nathan Dupont, Felix A. Palm and Amit Vashisht for useful discussions.}
This work is supported by the Hainan Province Science and Technology Talent Innovation Project (Grant No.~KJRC2023L05). XL acknowledges the Science Research Project of Anhui Educational Committee (2023AH050073) and the University Synergy Innovation Program of Anhui Province under Grant No. GXXT-2022-039.
NG and BW acknowledge the financial support from the ERC (LATIS project), the
EOS project CHEQS, the FRS-FNRS Belgium and the Fondation ULB.
\end{acknowledgments}	

%

\appendix

\section{GRAPE for Harvard experiment}\label{appendix:GRAPE}

\begin{figure}
	\centering
	\includegraphics[width=0.48\linewidth, trim= 100 230 100 250,clip]{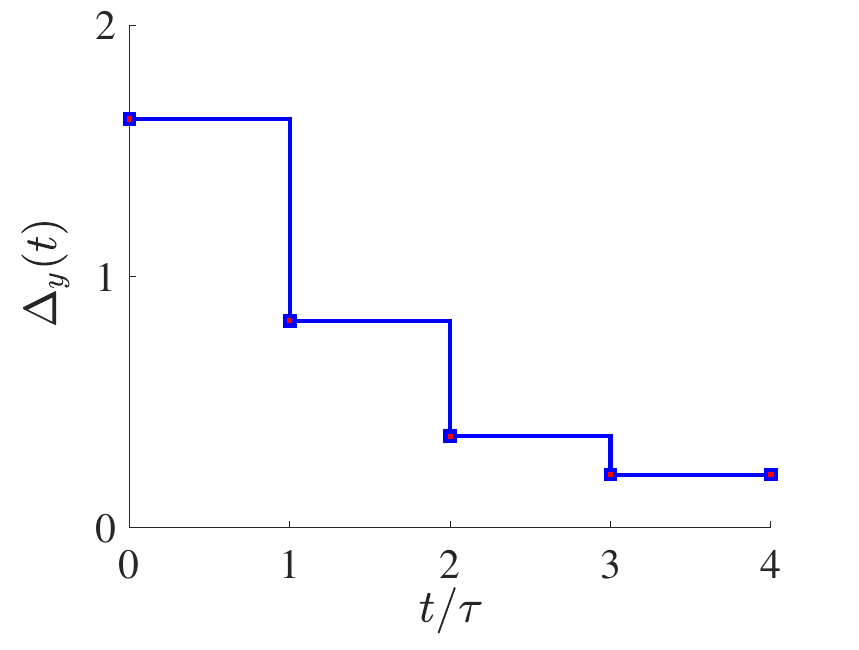}
    \includegraphics[width=0.48\linewidth, trim= 100 240 100 250,clip]{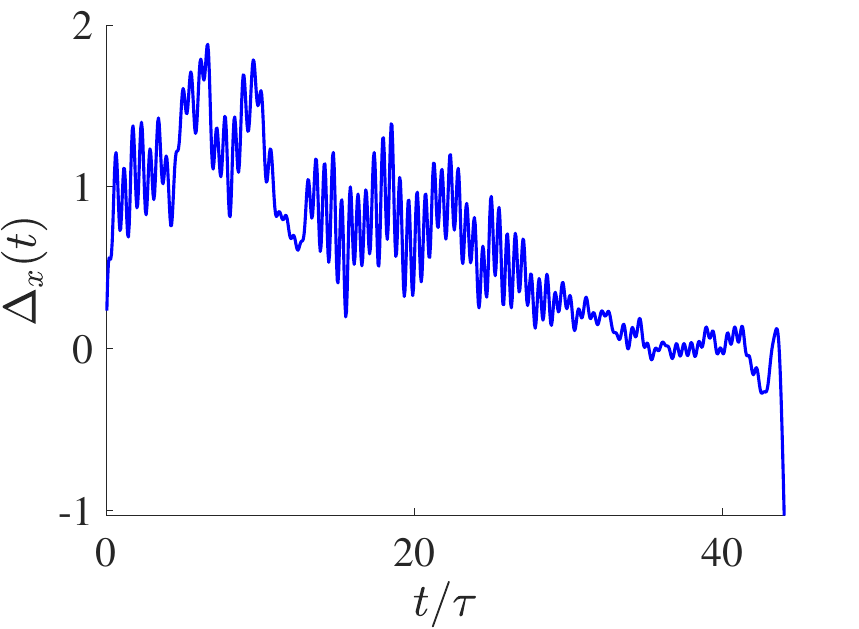}\\
	\caption{Control fields $\Delta_y(t)$ and $\Delta_x(t)$ optimized using GRAPE. Left: Optimized control field $\Delta_y(t)$ in step 2 in scheme 1. A nearly perfect fidelity $F\!=\!99.3\%$ is obtained with a short duration $T_2 \!=\! 4 \tau$. Right: Optimized control field $\Delta_x(t)$ in step 4. A nearly perfect fidelity $F\!=\!99.1\%$, with the Laughlin-type FQHE state, is obtained with duration $T_4 \!=\! 44 \tau$.
	}\label{fig:GRAPE}
\end{figure}

To benchmark the results obtained using smooth control protocol, We employ 
GRadient Ascend Pulse Engineering (GRAPE)~\cite{KHANEJA2005} method to optimize piecewise constant control fields, for the $4 \times 4$ lattice system. The four-step-ramping protocol is as follows:

In step 1, we take $T_1\!=\!6 \tau$ to linearly increase the tunneling $t_y$ from 0 to $\hbar/\tau$. In step 2 the minimal duration is estimated $T_2\!=\!4 \tau$, as well as the optimized control field $\Delta_y(t)$ by using GRAPE (see left panel of Fig.~\ref{fig:GRAPE}). At the end of step 2, the two particles are delocalized into one column. The final state of time evolution in step 2 is the initial state of step 3.

In step 3, we linearly increase the tunneling strength $t_x$ from 0 to $\hbar/\tau$ over time $T_3\!=\! 5 \tau$. In step 4, a high fidelity $F\!=\!99.1\%$ is obtained with a duration $T_4\!=\!44\tau$ using GRAPE (see right panel of Fig.~\ref{fig:GRAPE}). The total duration of state preparation is $T\!=\!\sum_{n\!=\!1}^{4} T_n \!=\! 59\tau$, which is a bit longer than the minimal duration $T\!=\!54\tau$ using smooth control protocol (refer to Sec.~\ref{sec:scheme1}).

We notice that the optimized control field $\Delta_x(t)$ obtained with GRAPE oscillates rapidly  (see right panel of Fig.~\ref{fig:GRAPE}). This makes such control protocols less favorable in experiments. Such an issue could be resolved by adding a penalty function which penalizes rapidly fluctuating controls. However, it usually makes the optimization more complicated, and the minimal duration to reach the target state is in general longer than the one without penalty function. Therefore, we stick to the smooth function, as illustrated in the main text.

\section{Time correlated noise}~\label{app:OU}
We use Ornstein-Uhlenbeck (OU) process~\cite{gardiner2009stochastic} to model time-correlated noise.
It is defined by the stochastic differential equation (SDE):
\begin{equation}\label{SDE}
dX_t = \theta(\mu - X_t)dt + \sigma dW_t
\end{equation}
where $\theta > 0$ is the mean reversion rate, $\mu$ is the long-term mean, $\sigma$ is the volatility parameter, and $W_t$ denotes standard Brownian motion.
The solution to this SDE is:
\begin{equation}
X_t = X_0 e^{-\theta t} + \mu(1 - e^{-\theta t}) + \sigma \int_0^t e^{-\theta(t - s)} dW_s,
\end{equation}
This closed-form expression reveals three key components:
\begin{enumerate}
    \item An exponentially decaying initial condition $X_0 e^{-\theta t}$,
    \item A deterministic drift towards the mean $\mu(1 - e^{-\theta t})$,
    \item A stochastic integral term representing accumulated noise.
\end{enumerate}

The autocorrelation function of the OU process is derived as follows:
\begin{align}
\text{Cov}(X_t, X_{t+\tau}) &= \mathbb{E}[(X_t - \mu)(X_{t+\tau} - \mu)] \\ 
&= \frac{\sigma^2}{2\theta} e^{-\theta \tau}.
\end{align}
The correlation time $\tau_c = \frac{1}{\theta}$ quantifies the decay rate: after $\tau_c$, correlations drop to $1/e \approx 36.8\%$ of their initial value.

As $t \to \infty$, the OU process converges to a stationary Gaussian distribution:
\begin{equation}
X_t \sim \mathcal{N}\left(\mu, \frac{\sigma^2}{2\theta}\right).
\end{equation}
This equilibrium ensures that the process remains bounded yet fluctuates around $\mu$, mimicking realistic noise with memory.

The power spectral density (PSD) of the OU Process is given by
\begin{equation}\label{PSD}
    S(\omega) = \frac{\sigma^2}{\tau_c^{-2} + \omega^2}.
\end{equation}
It is a {Lorentzian function} (characteristic of mean-reverting processes). At low frequencies (\( \omega \ll \tau_c^{-1} \)), \( S(\omega) \approx {\sigma^2\tau_c^2} \) (constant).
 At high frequencies (\( \omega \gg \tau_c^{-1} \)), \( S(\omega) \approx \frac{\sigma^2}{\omega^2} \) (decays as \( \omega^{-2} \)).

To numerically simulate the OU process, we discretize the SDE~\eqref{SDE} into time steps $\Delta t$:
\begin{equation}
X_{t+\Delta t} = X_t + \theta(\mu - X_t)\Delta t + \sigma \cdot \Delta W_t,
\end{equation}
where $\Delta W_t \sim \mathcal{N}(0, \Delta t)$ is a Wiener increment sampled from a normal distribution with variance $\Delta t$.
This noise is incorporated into the ideal control fields in a manner analogous to how white noise is integrated, as elucidated in the main text.

\begin{figure}[!htp]
	\centering
    \includegraphics[width=0.99\columnwidth]{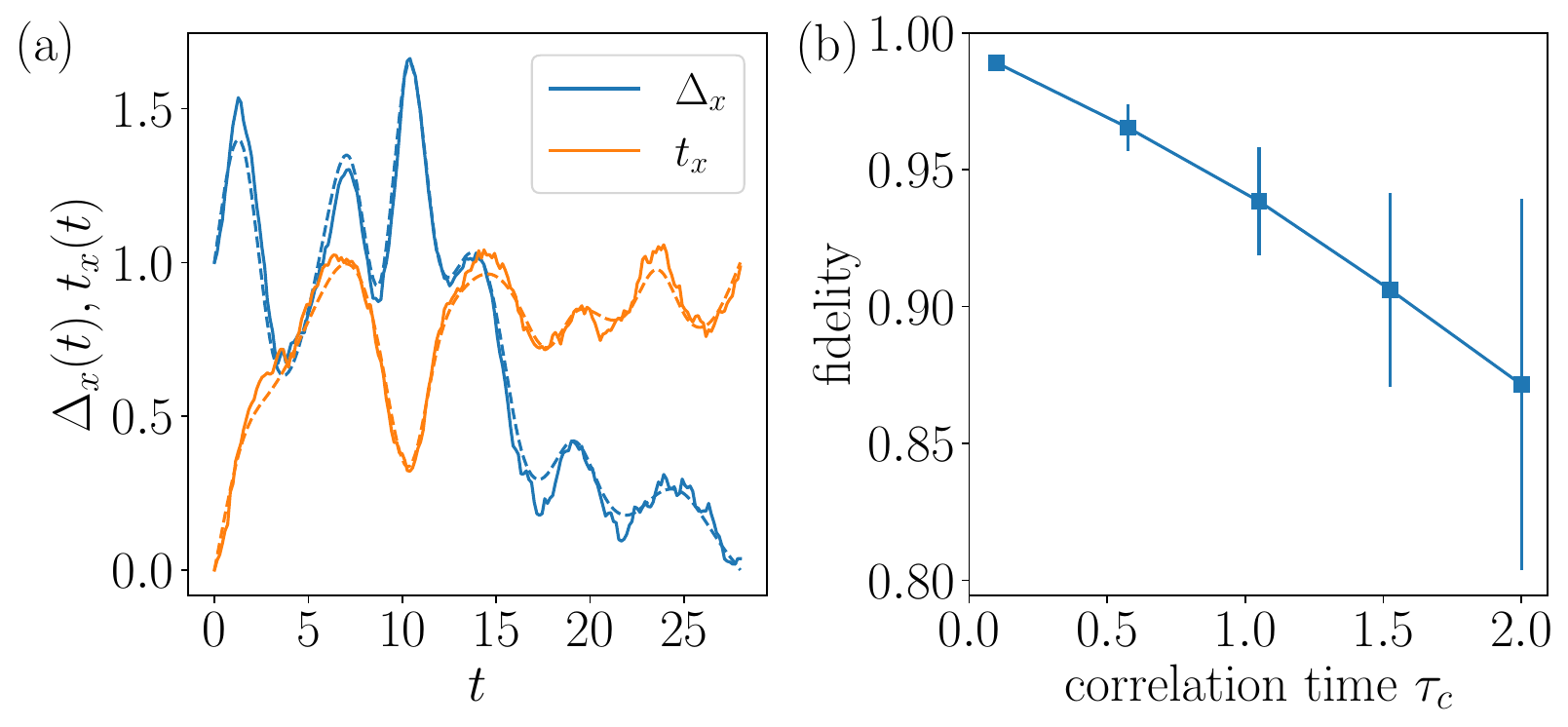}\\
	\caption{(a) Example noisy control profiles for Scheme II under the Ornstein-Uhlenbeck process  with correlation time $\tau_c=2\tau$. Dashed lines denote the ideal profiles.
    (b) Fidelity versus noise correlation time. 
   The results are obtained by averaging over $50$ runs. The errorbars denote one standard deviation. The parameters are $L_x\!=\!L_y\!=\!4$, $N\!=\!2$, $\phi\!=\!2\pi \times 0.26$, $U\!=\!8\hbar/\tau$. }\label{OU}
\end{figure}

The results are presented in Fig.~\ref{OU}, where panel (a) shows example noisy control profiles under the OU process  with correlation time $\tau_c=2\tau$,
and panel (b) quantifies the fidelity as a function of noise correlation time $\tau_c$. The results demonstrate a clear monotonic decrease in fidelity with increasing $\tau_c$, 
indicating reduced performance against more strongly correlated noise. 
This trend can be understood by examining the frequency-domain characteristics of the OU noise: its PSD~\eqref{PSD} shows that longer correlation times shift the noise power toward lower frequencies. 
For large $\tau_c$ values (corresponding to slowly varying noise), the spectral weight becomes concentrated in the low-frequency regime. 
Hence, these results align with our frequency-dependent analysis of white noise in Fig. ~\ref{noise4}(d) of the main text.
In both cases, we observe the same trend---our protocol demonstrates greater robustness to higher frequency noise components.

\end{document}